\documentstyle[aps,floats,epsf,psfig,twocolumn]{revtex}
\tighten
\begin{document}
\draft 
\title{Peierls Dimerization with Non--Adiabatic Spin--Phonon Coupling} 
\author{G. Wellein$^{1}$, H. Fehske$^{1}$, and A.P. Kampf$^{2}$}
\address{$^{1}$
Physikalisches Institut, Universit\"at Bayreuth, 95440 Bayreuth, Germany}
\address{$^{2}$
Institut f\"ur Physik, Theoretische Physik III, Elektronische Korrelationen 
und Magnetismus,\\
Universit\"at Augsburg, 86135 Augsburg, Germany}
\address{~
\parbox{14cm}{\rm 
\medskip
We study the magnetic properties of a frustrated Heisenberg spin chain with a 
dynamic spin--phonon interaction. By Lanczos diagonalization, preserving the 
full lattice dynamics, we explore the non--adiabatic regime with phonon 
frequencies comparable to the exchange coupling energy which is e.g. the 
relevant limit for the spin--Peierls compound $\rm CuGeO_3$. When compared to 
the static limit of an alternating spin chain the magnetic properties are 
strongly renormalized due to the coupled dynamics of spin and lattice degrees 
of freedom. The magnitude of the spin triplet excitation gap changes from a 
strong to a weak dimerization dependence with increasing phonon frequencies 
implying the necessity to include dynamic effects in an attempt for a 
quantitative description of the spin--Peierls state in $\rm CuGeO_3$.
 }}
\maketitle

\narrowtext
Low dimensional quantum spin systems have attracted considerable attention of 
theorists over the decades. Most of the remarkable features observed in these 
systems are pure quantum effects uniquely due to their low dimensionality. 
Among the current experimental and theoretical efforts for understanding the 
magnetic properties of linear chain and ladder materials the recent discovery 
of a spin--Peierls (SP) transition at $T_{SP}=14.3K$ in the inorganic compound
$\rm CuGeO_{3}$ (CGO) \cite{Hase} has received particular attention. In analogy
to the Peierls instability towards dimerization in quasi one-dimensional metals
\cite{Peierls} the energy of spin chains is lowered by dimerizing into an 
alternating pattern of weak and strong bonds. The SP transition is therefore 
driven by the magnetic energy gain which overcompensates the lattice 
deformation energy \cite{Cross}. In the SP phase of CGO the copper moments form
singlet dimers along the chains and there is an energy gap to spin triplet 
excitations. Experimentally, the SP nature of the transition and the spin gap 
have been firmly established by inelastic neutron scattering (INS), 
susceptibility, $X$--ray, and electron--diffraction experiments 
\cite{Regnaultrev}.

The Peierls transition naturally involves also the lattice degrees of freedom 
due to the alternating distortion of atomic positions. A rather common 
situation has been the clear separation of the electronic or magnetic energy 
scales from the frequencies of those phonons which couple most effectively to
the SP lattice distortion. This adiabatic limit is e.g. realized in 
trans--polyacethylene \cite{poly} or organic SP materials \cite{Bray}. In 
contrast, in CGO two weakly dispersive optical phonon modes involved in the SP 
transition have been identified with frequencies $\hbar\omega\approx J$ and 
$\hbar\omega\approx 2J$, where $J$ is the superexchange coupling between 
neighboring $\rm Cu$ spins \cite{Braden}. These hard phonons require a very 
strong spin--phonon coupling for a SP transition to occur \cite{Bulaevski}. A 
strong magnetoelastic coupling has indeed been observed in thermal expansion 
and magnetostriction experiments \cite{BerndPRL} and inferred from large 
uniaxial pressure ($p_i$) derivatives $\partial J/\partial p_i$ 
\cite{Fabricius}. The SP physics in CGO is therefore in a non--adiabatic 
regime. 

Despite this unusual situation previous theoretical studies have commonly 
adopted an alternating and frustrated antiferromagnetic (AF) Heisenberg spin 
chain model \cite{Castilla}
\begin{equation}
H=J\sum_{i}\left[(1+\delta(-1)^i)\,{\bf S}_i\cdot{\bf S}_{i+1}+\alpha\,
{\bf S}_i\cdot{\bf S}_{i+2}\right]
\label{hstatic}
\end{equation}
with a static dimerization parameter $\delta$, thus representing the extreme 
adiabatic limit of a SP chain. $i$ denotes the sites of a chain with length 
$N$ and ${\bf S}_i$ are $S=1/2$ spin operators. $\alpha$ determines the 
strength of a frustrating AF next--nearest neighbor coupling. The 
model~(\ref{hstatic}) contains two independent mechanisms 
for spin gap formation. At $\delta=0$ and for $\alpha<\alpha_c$ the groundstate
is a spin liquid and the elementary excitations are massless spinons 
\cite{Haldane1}. $\alpha_c=0.241$ was accurately determined by numerical 
studies \cite{Okamoto,Castilla}. For $\alpha>\alpha_c$ the groundstate is 
spontaneously dimerized, the spectrum acquires a gap and the elementary 
excitations are massive spinons \cite{Chitra,White}. On the other hand for any 
finite $\delta$ the singlet groundstate of (\ref{hstatic}) is also
dimerized, but the elementary excitation is a massive magnon 
\cite{Haldane1,Tsvelik}. 

From fits to magnetic susceptibility data in the uniform high temperature 
phase \cite{Riera,Fabricius} and from the requirement to reproduce the 
experimental triplet excitation gap \cite{Riera,Schoenfeld} the parameter set 
$J=160K$, $\alpha=0.36$, and $\delta=0.014$ was estimated for CGO within the 
static model~(\ref{hstatic}). Yet, the amplitude of the dimerization is 
substantially underestimated when compared to estimates from structural data in
the SP phase \cite{Berndunp,Braden}. In this letter we will show that due to 
the non--adiabatic phonon dynamics a much larger dimerization is needed to 
achieve the same magnitude of the triplet gap and thus offers a possible route 
towards a quantitatively consistent description. 

The dimerization parameter $\delta$ in Eq.~(\ref{hstatic}) may be viewed to
result from a mean-field treatment of an interchain coupling which is generated
by a purely  elastic mechanism \cite{Essler}. In fact, this coupling to 
neighboring dimerized chains provides an external potential lifting the 
degeneracy between the two possible dimerized groundstates \cite{Khomskii2}.
Thereby, soliton and antisoliton excitations form spin triplet or spin singlet 
bound states \cite{Georges}, the former is the elementary massive magnon of 
the spin Hamiltonian~(\ref{hstatic}) for $\delta>0$.

\begin{figure}
\mbox{\epsfxsize 8.5cm\epsffile{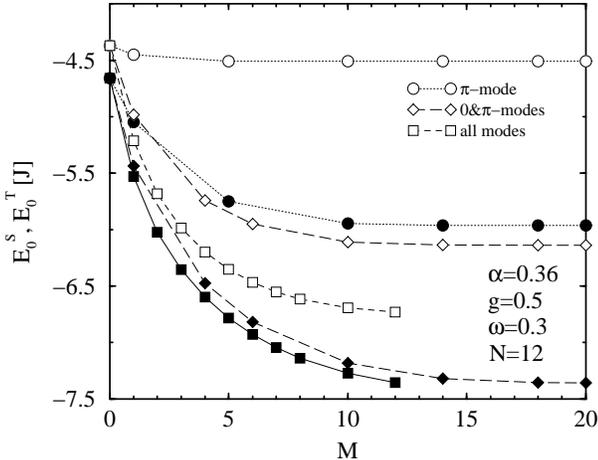}}
\caption[]{
Energies of the singlet groundstate (full symbols) and the lowest triplet 
excitation (open symbols) vs. the number of phonons used in the numerical 
diagonalization. The different symbols indicate the results obtained by 
including phonons (a) only with $q=\pi$ ($\circ$), (b) with $q=0$ and 
$q=\pi$ ($\Diamond$), and (c) at all wavevectors ($\Box$). As in all subsequent
figures energies are measured in units of J. }
\label{fig1}
\end{figure} 

In this letter we choose a different starting point and use a Heisenberg chain
model in which the spins are explicitly coupled to the lattice degrees of 
freedom maintaining their full quantum dynamics. Specifically we study the 
magnetic properties of the Hamiltonian \cite{Khomskii2}
\begin{eqnarray}
H_{SP}&=&J\sum_i\left[\left(1+g(b_i^++b_i)\right){\bf S}_i\cdot{\bf S}_{i+1}+
\alpha\,{\bf S}_i\cdot{\bf S}_{i+2}\right] \nonumber \\
&+&\hbar\omega\sum_ib^+_ib_i 
\label{hsp}
\end{eqnarray}
with the local phonon creation and annihilation operators $b_i^+$ and $b_i$,
respectively. For the optical phonon with frequency $\omega$ we thus choose
a dispersionless Einstein mode -- an appropriate choice for modeling those 
phonons in CGO which are most relevant for the Peierls distortion 
\cite{Braden}. In the non--adiabatic ($\hbar\omega\sim J$) and intermediate 
coupling ($g\sim J$) regime the spin and the phonon dynamics are intimately 
coupled and integrating out the phonons in order to obtain an effective 
renormalized spin--only Hamiltonian is in general not possible \cite{Uhrig}.  
 
This situation thus requires a numerical approach. We have performed exact
diagonalizations on finite chains up to $N=16$ sites with periodic boundary 
conditions. Since the Hilbert space associated with the phonons is infinite 
even for a finite system, we apply a controlled truncation procedure retaining
only basis states with at most $M$ phonons \cite{Baeuml}. $M$ is fixed by
requiring a relative error in the groundstate energy less than ${10}^{-7}$. In
our calculations all possible phonon modes are taken into account. 

Fig.~\ref{fig1} demonstrates for a 12 sites chain how many phonons have to be 
included to achieve convergence for the groundstate energies of the singlet 
$S=0$ and the triplet $S=1$ sectors. Although the specific value of $\alpha$ is
not crucial for our arguments, we keep the frustration parameter $\alpha=0.36$ 
in all calculations fixed to its most likely value in CGO. Fig.~\ref{fig1} 
contrasts the results obtained if only phonons with wavevector $q=\pi$ or 
$q=\pi$ {\it and} $q=0$ are used to the case when all phonon modes are 
included. While it has been argued in Ref. \cite{AugierPoilblanc} that keeping 
the $q=\pi$ mode only should already capture the dominant dimerization effect 
of the spin--phonon interaction, we find in Fig.~\ref{fig1} that e.g. the 
singlet--triplet excitation gap $\Delta^{ST}=E_0^T-E_0^S$ is strongly 
renormalized when phonons of {\it all} $q$ are taken into account implying that
a spin triplet excitation is accompanied by a {\it local} distortion in the 
lattice. 

\begin{figure}
\mbox{\epsfxsize 8.5cm\epsffile{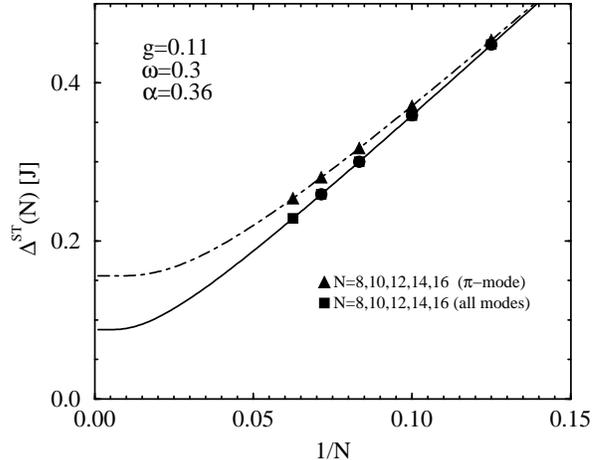}}
\caption[]{
Finite size scaling behavior of the triplet excitation gap $\Delta^{ST}$ for  
$\alpha=0.36$. The lines result from the fit to the scaling 
function~(\ref{scaling}). The (almost adiabatic) phonon parameters $g=0.11$ 
and $\omega=0.3$ were used in order to make contact with previous numerical 
studies \cite{AugierPoilblanc}.}
\label{fig2}
\end{figure} 

The renormalization of the singlet--triplet gap $\Delta^{ST}$ is shown in the 
scaling plot in Fig.~\ref{fig2}. It is contrasted to the result with a 
restriction to the $q=\pi$ phonons only. Despite of the limited system sizes 
of up to 16 sites the data are well fitted by the scaling function 
\cite{Barnes,Schoenfeld}
\begin{equation}
\Delta^{ST}(N)=\Delta^{ST}+{A\over N}e^{-N/N_0}\,\, . 
\label{scaling}
\end{equation}
The data clearly demonstrate that the restriction to the $\pi$--modes 
substantially overestimates the size of the gap for $N\rightarrow\infty$.

\begin{figure}
\mbox{\epsfxsize 8.5cm\epsffile{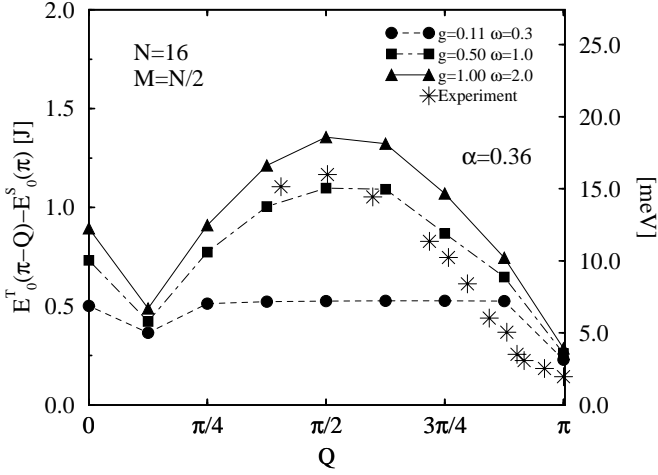}}
\caption[]{
Dispersion of the elementary triplet excitation (left scale) for a chain with 
$N=16$ sites for different phonon frequencies and spin--phonon coupling 
constants. Experimental data (*, right scale) are also included from Ref. 
\cite{Regnault}.}
\label{fig3}
\end{figure}

The triplet dispersion shown in Fig.~\ref{fig3} supports the expectation -- as 
inferred from the neutron scattering studies of the phonons \cite{Braden} -- 
that the frequencies of the relevant optical phonon modes in CGO are comparable
to or larger than the exchange coupling $J$. Because for $\hbar\omega<J$ the 
phononic character of the excitation dominates leading to a flattening of the 
dispersion \cite{comment,wefe}. This, however, is incompatible with the 
magnetic INS data \cite{Regnault} which do not show any signature for a flat 
dispersion of the triplet excitation branch. We have included in 
Fig.~\ref{fig3} experimental data on CGO from Ref. \cite{Regnault} which, 
however, is not intended as a fit to the data -- our results in Fig.~\ref{fig3}
are for a fixed chain length with $N=16$ sites -- but rather shows the shape of
the triplet dispersion. We stress that the upturn at $Q=0$ is a finite--size 
effect which also occurs for the pure spin model and rapidly decreases with 
increasing $N$ \cite{WelleinPhD}; for $N\rightarrow\infty$, $E^T_0(\pi)$ and 
$E^T_0(0)$ become degenerate.
 
\begin{figure}
\mbox{\epsfxsize 8.5cm\epsffile{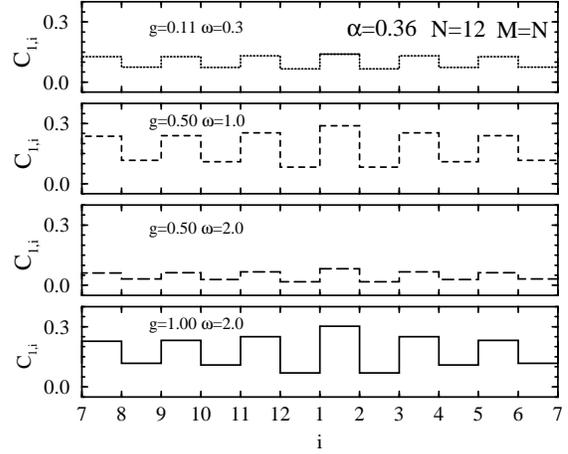}}
\caption[]{
Displacement correlation function $C_{1,i}$ as defined in the text for 
different phonon frequencies and spin lattice coupling constants for a chain 
with $N=12$ sites.}
\label{fig4}
\end{figure}

The Peierls ordering structure of the groundstate can be demonstrated in two 
complementary ways. The {\it magnetic} ordering pattern of an alternating local
singlet formation is captured by the order parameter
\begin{equation} 
D=\langle{1\over N}\sum_i(-1)^i\left({\bf S}_{i-1}\cdot{\bf S}_i-{\bf S}_i
\cdot{\bf S}_{i+1}\right)\rangle\,.
\label{corrsp}
\end{equation}
$D$ is the proper choice for the Heisenberg spin chain 
model~(\ref{hstatic}). For the translationally invariant quantum spin--phonon 
Hamiltonian~(\ref{hsp}) the Peierls distortion of the lattice is reflected in 
the spatial structure of the displacement correlation function 
\begin{equation}
C_{1,i}=\langle\left(b_1+b_1^+\right)\left(b_i+b_i^+\right)\rangle-\delta_{1,i}
\,\, . 
\label{correlu}
\end{equation}
The alternating structure of the correlation function $C_{1,i}$ shown in 
Fig.~\ref{fig4} implies the Peierls formation of short and long bonds and thus
alternating strong and weak AF exchange interactions, i.e. a dimerized 
groundstate. Naturally the dimerization is enhanced (weakened) by increasing 
the spin--phonon coupling (phonon frequency). We note that for any finite 
phonon frequency there exists a critical coupling constant $g_c$ 
\cite{PoilblancAffleck}; in the infinite system quantum lattice fluctuations 
will destroy the Peierls dimerization for couplings $g<g_c$. For non--adiabatic
phonon frequencies $\hbar\omega\sim J$ also the critical coupling $g_c$ is 
comparable to $J$ -- a situation which matches the strong magnetoelastic 
coupling and the SP physics in CGO.

\begin{figure}
\mbox{\epsfxsize 8.5cm\epsffile{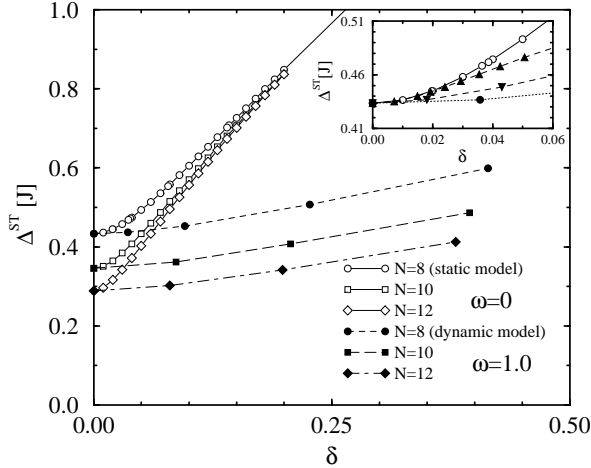}}
\caption[]{
Triplet excitation gap $\Delta^{ST}$ versus dimerization $\delta$ for the 
static spin--only ($\omega=0$, $\diamond\,\Box\,\circ$) and the dynamic 
spin--phonon model (full symbols) with phonon frequency $\omega=1.$ for 
different chain lengths at $\alpha=0.36$. The inset shows the results for the 
dynamic eight--site model (filled symbols; $\omega=0.1$ [$0.3$] triangles up 
[down], 1.0 ($\bullet$)) in comparison to the static case ($\circ$).}
\label{fig5}
\end{figure}

In order to quantify these results we obtain the magnitude of the Peierls 
dimerization order parameter $\delta$ from the static (lattice) structure 
factor at wavenumber $q=\pi$ 
\begin{equation}
\delta^2=\frac{g^2}{N^2}\sum_{i,j}\langle u_iu_j\rangle
\mbox{e}^{{\rm i}\pi(R_i-R_j)} 
\label{deltaq}
\end{equation} 
where $u_i=b_i+b_i^+$. Like in the ordinary Peierls phenomenon, a finite 
dimerization $\delta>0$ necessarily leads to a gap $\Delta^{ST}$ in the 
magnetic excitation spectrum. For evaluating the relation between the 
dimerization and the resulting magnitude of the spin triplet excitation gap we 
keep the phonon frequency fixed, vary the coupling strength $g$ and calculate 
for each parameter set $\Delta^{ST}$ (see Fig.~\ref{fig5}), and the 
dimerization $\delta$ from Eq. (\ref{deltaq}). 

The results for the static (spin--only) and the dynamic spin--phonon model are 
compared in Fig.~\ref{fig5} for three different chain lengths. For vanishing 
lattice dimerization, i.e. in the absence of any spin--phonon coupling $(g=0)$,
the results for the static and the dynamic model naturally agree (note that for
$\delta=0$ there remains a spin excitation gap even for $N\rightarrow\infty$ 
due to the frustration driven singlet dimer ordering $D\neq 0$). The important 
message of Fig.~\ref{fig5} is that a substantial enhancement of the 
dimerization is needed to achieve a given triplet excitation gap, if the
dynamic phonons have frequencies comparable to $J$. This partially resolves the
quantitative problem encountered within the alternating and frustrated 
Heisenberg spin chain model~(\ref{hstatic}), because fixing parameters $J$ and 
$\alpha$ by fitting the high temperature susceptibility and $\delta$ by 
matching $\Delta^{ST}$ to the INS data in CGO, the dimerization parameter 
$\delta$ is underestimated by roughly a factor of 5~\cite{Berndunp,brapriv}. 
The inset in Fig.~\ref{fig5} shows the evolution of the $\Delta^{ST}$ vs. 
$\delta$ results with increasing phonon frequency $\omega$ at a fixed chain 
length. With decreasing $\omega$ the data smoothly merge with the result for 
the static spin--only model.

In conclusion, we have numerically analyzed the Peierls dimerized groundstate 
and the spin triplet excitations in a Heisenberg chain model with dynamic
spin--phonon coupling. The magnetic excitations inherently include a local 
lattice distortion requiring a multi--phonon mode treatment of the lattice 
degrees of freedom. A spin--phonon coupling induced flattening of the triplet 
dispersion is avoided for a strong spin--phonon coupling in the non--adiabatic 
regime. The non--adiabatic phonon dynamics strongly renormalizes the magnetic 
excitation spectrum and the dimerization dependence of the triplet excitation 
gap. These features are in accordance with experimental data on CGO suggesting 
the necessity for a non--adiabatic spin--phonon approach to the SP physics in 
this material. 

We acknowledge useful discussions with G. Bouzerar, M. Braden, D. Khomskii, W. 
Weber, and A. Wei{\ss}e. We are particularly indebted to B. B\"uchner for 
informing us about unpublished estimates for the dimerization of $J$ in the SP
phase of CGO. The numerical calculations were performed on the 
vector--parallel supercomputer Fujitsu VPP700 at the LRZ M\"unchen.


\begin{references}
\bibitem{comment} We emphasize that the flattening of the dispersion can not be
obtained if only phonon modes with $q=\pi$ are included as in 
Ref.~\cite{AugierPoilblanc}.
\end{references}
\end{document}